\documentclass[twocolumn,twocolappendix]{aastex63}

\def\etal{et~al.\ }

\usepackage{braket}

\usepackage[utf8]{inputenc}
\usepackage{CJK}
\usepackage{enumitem}

\received{}
\revised{}
\accepted{}
\submitjournal{ApJ}

\shorttitle{Cosmological Information from the Small-scale RSD}
\shortauthors{Tonegawa et al.}
\begin{document}

\title{Cosmological Information from the Small-scale Redshift Space Distortions}

\begin{CJK*}{UTF8}{mj}

\author{Motonari Tonegawa}
\affiliation{School of Physics, Korea Institute for Advanced Study,
85 Hoegiro, Dongdaemun-gu, Seoul, 02455, Korea}
\author[0000-0001-9521-6397]{Changbom Park}
\affiliation{School of Physics, Korea Institute for Advanced Study,
85 Hoegiro, Dongdaemun-gu, Seoul, 02455, Korea}
\author{Yi Zheng}
\affiliation{School of Physics and Astronomy, Sun Yat-sen University, 2 Daxue Road, Tangjia, Zhuhai, 519082, China}
\author[0000-0002-7464-7857]{Hyunbae Park (박현배)}
\affiliation{School of Physics, Korea Institute for Advanced Study,
85 Hoegiro, Dongdaemun-gu, Seoul, 02455, Korea}
\affiliation{Korea Astronomy and Space Science Institute, 776 Daedeok-daero, Yuseong-gu, Daejeon, 34055, Korea}
\affiliation{Kavli IPMU (WPI), UTIAS, The University of Tokyo, Kashiwa, Chiba 277-8583, Japan}
\author[0000-0003-4923-8485]{Sungwook E. Hong (홍성욱)}
\affiliation{Natural Science Research Institute, University of Seoul,
163 Seoulsiripdaero, Dongdaemun-gu, Seoul, 02504, Korea}
\author[0000-0003-3428-7612]{Ho Seong Hwang}
\affiliation{Korea Astronomy and Space Science Institute, 776 Daedeok-daero, Yuseong-gu, Daejeon, 34055, Korea}
\author[0000-0002-4391-2275]{Juhan Kim (김주한)}
\affiliation{Center for Advanced Computation, Korea Institute for Advanced Study,
85 Hoegiro, Dongdaemun-gu, Seoul, 02455, Korea}

\correspondingauthor{Juhan Kim} \email{kjhan@kias.re.kr}



\begin{abstract}
The redshift-space distortion (RSD) in the observed distribution
of galaxies is known as a powerful probe of cosmology. Observations of large-scale RSD, caused by the coherent gravitational infall of galaxies, have given tight constraints on the linear growth rate of the large-scale structures in the universe.
On the other hand, the small-scale RSD, caused by galaxy random motions inside clusters,
has not been much used in cosmology, but also has cosmological information because universes with different cosmological parameters have different halo mass functions and virialized velocities.
We focus on the projected correlation function $w(r_p)$ and the multipole moments $\xi_l$ on small scales ($1.4$ to $30\ h^{-1}\rm{Mpc}$).
Using simulated galaxy samples generated from a physically motivated \replaced{halo-galaxy}{most bound particle (MBP)-galaxy} correspondence scheme in the Multiverse Simulation, we examine the dependence of the small-scale RSD on the cosmological matter density parameter $\Omega_m$, the satellite velocity bias with respect to MBPs, $b_v^s$, and the \replaced{merger time scale}{merger-time-scale parameter} $\alpha$.
We find that $\alpha=1.5$ gives an excellent fit to the $w(r_p)$ and $\xi_l$ measured from the SDSS-KIAS value added galaxy catalog.
We also define the ``strength'' of Fingers-of-God as the ratio of the parallel and perpendicular size of the contour in the two-point correlation function set by a specific threshold value and show that the strength parameter helps constraining $(\Omega_m, b_v^s, \alpha)$ by breaking the degeneracy among them.
The resulting parameter values from all measurements are $(\Omega_m,b_v^s)=(0.272\pm0.013,0.982\pm0.040)$,
indicating a slight reduction of satellite galaxy velocity relative to the MBP.
However, considering that the average MBP speed inside haloes is $0.94$ times the dark matter velocity dispersion,
the main drivers behind the galaxy velocity bias are gravitational interactions, rather than baryonic effects.
\end{abstract}


\section{Introduction}
The accelerated expansion of the Universe has been one of the most profound mysteries
in astronomy and physics since observations have confirmed it through the redshift-distance relation of type Ia supernovae (\citealt{Riess1998,Perlmutter1999}).
%and the temperature fluctuation of the cosmic microwave background (\citealt{Spergel2007}).
So far, the $\Lambda$CDM model gives the best description for these observations,
although it involves several theoretical difficulties related to the smallness and fine-tuning of $\Lambda$, the exotic form of the energy in the Universe  (\citealt{Frieman2008,Weinberg2013}).
Another conceptual possibility for the apparent accelerated expansion is that
the general relativity (GR), on which the $\Lambda$CDM model is built,
may not be correct in cosmological scales. 
This idea gave rise to the modified gravity theories,
which realized the same redshift-distance relation as that of $\Lambda$CDM model
without relying on the dark energy
but predict different gravitational growth history of the matter content of the Universe (\citealt{Joyce2016,Koyama2016}).
Discriminating between the dark energy and modified gravity scenarios is essential to understand the origin and history of our Universe better.

\replaced{Redshift}{The redshift} space distortion (RSD) is a phenomenon 
that the \replaced{observationally inferred}{observed} distribution of galaxies is distorted from the \replaced{true}{real} one due to \replaced{the shift in observed redshifts}{the non-cosmological redshift}
caused by galaxy peculiar motion \citep{Kaiser1987, Hamilton1998}. %% Jackson1972?
It affects the statistical property of galaxy clustering such as the two-point correlation function (2pCF) and the power spectrum, making the line-of-sight direction a special one.
As the galaxy peculiar velocity field is governed by the gravity law and background cosmological parameters, the anisotropy of the galaxy 2pCF is sensitive to
the change of cosmological models, making RSD a powerful cosmological probe \citep{Weinberg2013}.
Since the galaxy catalog used in an RSD analysis can also be used for \replaced{other geometrical tests of the Universe}{other cosmological probes}
such as large-scale structure topology \citep{Park2010,Appleby2018}, richness and size distributions of structures \citep{Hwang2016}, baryon acoustic oscillations (BAO), and Alcock-Paczynski test \citep{Reid2012, Li2016, Sanchez2017},
there have been a variety of galaxy redshift surveys (SDSS: \citealt{York2000};
HectoMAP: \citealt{Geller2011}; BOSS: \citealt{Dawson2013}; 6dF: \citealt{Jones2005}; WiggleZ: \citealt{Drinkwater2010}; VIPERS: \citealt{Guzzo2014}; FastSound: \citealt{Tonegawa2015}; eBOSS: \citealt{Dawson2016}). 
There are also further large upcoming surveys (PFS: \citealt{Takada2014}; DESI: \citealt{DESI2016}; WFIRST: \citealt{Spergel2015}).

The large-scale RSD is caused by the infall motion of galaxies during the structure formation, and it has been detected by various redshift surveys,
giving strong cosmological constraints on the growth rate of the large-scale structure $f=d\ln{D}/d\ln{a}$ \citep{Hawkins2003, Guzzo2008, Blake2011, Beutler2012, Samushia2012, delaTorre2013, Beutler2014, Okumura2016, Icaza-Lizaola2019}, where $D$ is the growth factor and $a$ is the scale factor of the universe
with $a=1$ at the present epoch.
On the other hand, the small-scale RSD, called finger-of-god (FoG) effect,
is caused by the orbital motion of galaxies inside galaxy groups and clusters.
It has not been as much studied as the large-scale RSD but has rich cosmological information because different cosmological parameters lead to different halo mass functions and virialized velocities \citep{Marzke1995}.
The difficulty in using the small-scale RSD lies in the theoretical prediction
of the density and velocity field in highly non-linear scales.
We cannot rely on the perturbation theory that is valid down to mildly non-linear regime \citep{Taruya2010},
because the FoG effect takes place in the almost or completely relaxed objects.
There \replaced{are}{have been} attempts to \replaced{model the small-scale redshift space 2pCF by modeling
the pairwise velocity of galaxies}{understand the pairwise velocity of galaxies, which is an essential ingredient for the small-scale redshift space 2pCF} (\citealt{Sheth1996,Juszkiewicz1998};  \citealt{Tinker2007}; \citealt{Bianchi2016}; \citealt{Kuruvilla2018}).
\added{While they gave illuminating insights on why the pairwise velocity distribution has the shape of what we observe,}
their models typically include free parameters which may depend on cosmological models and are not easy to derive from the first principles thus far.
%it is difficult to model RSD \delted{over all scales} from the first principles.
Nevertheless, because of the small statistical uncertainty,
the use of the small-scale clustering will significantly enhance
our constraining power from the limited size of observational data sets.

While the analytic prescription of the non-linear structure formation is a timely topic itself (\citealt{Tinker2007}),
$N$-body simulations can serve as an alternative \replaced{means for}{to} small-scale cosmology studies (\citealt{DeRose2019}).
In this study, we use the Multiverse \replaced{simulation \citep{Kim2015}}{Simulation (\citealt{Shin2017, Park2019, Hong2020} for scientific applications)}, the Horizon Run 4 Simulation \citep{Kim2015}, and
a physically motivated galaxy assigning scheme \citep{Hong2016}
to \replaced{simulate}{mock} the galaxy distributions in redshift space for different matter density parameters $\Omega_m$, satellite velocity bias parameters $b_v^s$, and merger time scale parameters, $\alpha$. 
The galaxy-halo correspondence in our model has more physical meaning than
the halo occupation distribution (HOD) approach.
\citet{Reid2014} adopted the HOD approach and successfully explained the redshift-space clustering of the BOSS CMASS data, obtaining a $2.5\%$ constraint of the growth rate, which clearly proved the usefulness the small-scale clustering information.
Although being the standard method to connect galaxies and haloes,
the HOD has several issues to examine carefully.
The HOD prescribes the probability of a halo of mass $M$ having $N$ galaxies, $P(N|M)$, with typically five parameters and specific functional forms.
However, there is not a particular reason for the number of parameters and the functions. The number $N$ may also depend on secondary parameters, such as halo age and galaxy assembly history \citep{Wang2013,Dorta2017,Beltz-Mohrmann2019}.
By contrast, our galaxy-halo corresponding scheme traces the merger tree and automatically places galaxies into subhaloes, which avoids the theoretical uncertainties.
We constrain the matter density parameter $\Omega_m$ as well as the velocity bias parameter for satellite galaxies $b_v^s$ and the merger time scale parameter $\alpha$ by simultaneously matching the measurements of the projected correlation function and the multipole moments of the two-point correlation function
%from the KIAS \replaced{value added galaxy catalog}{Value-Added Galaxy Catalog} (KIAS-VAGC; \citet{Choi2010}) data.
between simulation and observation.
We also define the ``strength'' of FoG and show that adding it helps us to constrain
our model parameters more strongly.

The structure of this paper is as follows.
In section \ref{section:data}, we describe the simulation and observational data that we use.
In section \ref{section:measurements}, we measure the correlation functions and covariance matrix. We also quantify the FoG \replaced{sharpness}{strength} to extract cosmological information from the small-scale 2pCF.
In section \ref{section:results}, we show our constraints on the parameters of our model,
followed by discussions in section \ref{section:discussions}.
Finally we summarize our study in section \ref{section:conclusions}. 

\section{Data and Models}\label{section:data}
\subsection{The KIAS-VAGC catalog}
We use the KIAS \replaced{value added galaxy catalog (KIAS-VAGC)}{Value-Added Galaxy Catalog \citep[KIAS-VAGC;][]{Choi2010}} as observational data.
This catalog is based on the New York University Value-Added Galaxy Catalog (NYU-VAGC; \citealt{Blanton2005})
as part of Sloan Digital Sky Survey Data Release 7 \citep{Abazajian2009},
but supplements missing redshifts with other galaxy redshift catalogs for better redshift completeness.
The KIAS-VAGC covers $\sim 8000\ {\rm deg^2}$ on the celestial plane and contains $593,514$ redshifts of the SDSS Main galaxies in $r$-band Petrosian magnitude of $10.0<m_r<17.6$.
The supplementation increased the area with completeness higher than $0.97$,
from $39.8\%$ to $54.3\%$.
There are still missing redshifts even after this supplementation, which is mainly caused
by the fiber collision effect and poor observing conditions.
The fiber collision rate is estimated to be $\sim 5\%$, but lower in the overlapping regions. %%
In the KIAS-VAGC catalog, these galaxies are marked and given redshifts of the nearest galaxy on the celestial plane.

We use the volume-limited sample ``D5'' with a redshift cut $0.025<z<0.10713$ and an $r$-band absolute magnitude cut \replaced{${\cal M} < 20.02$}{$M_r < -20.02+5\log{h}$} as defined in \cite{Park2009}. 
The number density is $0.063 \ (h^{-1}\rm{Mpc})^{-3}$ and median redshift is $0.083$.
Also, we restrict the sample to the largest area that satisfies $-65.0^\circ<\lambda<65.0^\circ$ and $-37.0^\circ<\eta<43.0^\circ$, where $\lambda$ and $\eta$ are the SDSS survey coordinates.
The KIAS-VAGC also provides the survey mask, which indicates the spectroscopic completeness
in each of $0.025\times0.025 \ {\rm deg^2}$ patch in the survey area.
To avoid the bad observing condition and shot noise,
we only use the region where the completeness is above 0.8.
All galaxies in the valid region are assigned the weight as the inverse of the completeness.
Some of the target galaxies are not allocated fibers to, due to the mechanical limitation about
the minimum separation of two galaxies on the sky.
This is called the fiber collision effect, occurring on small scales ($\sim0.1\ {\rm Mpc}$)
and potentially weakens the FoG effect.
If a spectroscopic target cannot be allotted a fiber, the redshift of the nearest neighbor galaxy
is given to the galaxy. The validity of this approach will be discussed \replaced{later}{in appendix \ref{appendix:NNtest}}.

The random catalog is needed to measure the correlation function.
Since we are using a volume-limited sub-sample,
we make random catalogs as the uniform distribution in a comoving volume.
The angular completeness mask is then applied after the conversion from
$(X,Y,Z)$ to $(\lambda, \eta)$ to discard points on the region of ${\rm completeness}<0.8$.
The size of the random catalog is $\sim 30$ times larger than the corresponding data.

\subsection{The Multiverse Simulation}
The Multiverse \replaced{simulations are}{Simulation is} a collection of large\added{-}volume cosmological \added{$N$-body} simulations with different cosmological parameters \replaced{\citep{Kim2015}}{\citep{Shin2017, Park2019}}.
There are five realizations that have different matter density $\Omega_m$
and the equation of state of the dark energy $w$: $(\Omega_m, w)=(0.21,-1.0)$, $(0.26,-1.0)$, $(0.31,-1.0)$, $(0.26,-0.5)$ , and $(0.26,-1.5)$, keeping $\Omega_m+\Omega_\Lambda=1$,
while other parameters are fixed \replaced{as}{to} the Wilkinson Microwave Anisotropy Probe \added{(WMAP)} 5-year result \citep{Dunkley2009}: $\Omega_b=0.044$, $n=0.96$, $H_0=72 \ \rm{km\,s^{-1}\,Mpc^{-1}}$, and $\sigma_8=0.79$.
We use the first three \replaced{realizations}{simulated models} out of five 
because the change in $w$ will not be important for the clustering properties
on the scales that we are interested in. 
The comoving box size is $1024^3\  (h^{-1}\rm{Mpc})^3$ and $2048^3$ dark matter (DM) particles are evolved inside,
which leads to a particle mass of $9\times10^9  (\Omega_m/0.26)M_{\sun}$.
The starting redshift is $z=99$, and $1980$ snapshots are saved until $z=0$.
Haloes are identified through a friend-of-friend (FOF; \citealt{Davis1985}) algorithm 
with a commonly-used linking length of $b=0.2$.
The minimum number of particles to be qualified as a halo is $30$, which means the minimum halo mass of $2.7\times10^{11}(\Omega_m/0.26) M_{\sun} $.
Haloes in each snapshot are searched for the most bound particle (MBP), which is located at the lowest gravitational potential.
The merger tree is built by tracking the merger trajectories of \replaced{haloes}{MBPs}.

Simulated galaxies are assigned to the DM haloes by the MBP-galaxy correspondence approach as described by \citet{Hong2016}. 
All MBPs marked in the merger tree are regarded as galaxy proxies and physical properties of MBPs such as mass, position, and velocity are allocated to the modeled galaxies.
If a merger occurs, \replaced{the host and satellite MBPs are defined}{the roles of host and satellite are assigned to each MBP} according to the \replaced{size}{mass} of the haloes in the previous time step.
Then, the satellites are monitored to determine their fates (i.e., escape from the gravitational potential of its host or tidally disrupted), according to the \added{modified version of} merger time scale of \cite{Jiang2008}:
\begin{equation}\label{equation:Jiang08}
\frac{t_{\rm merge}}{t_{\rm dyn}}=\frac{(0.94\epsilon^{0.60}+0.60)/0.86}{\ln{[1+(M_{\rm host}/M_{\rm sat})]}} \left (\frac{M_{\rm host}}{M_{\rm sat}} \right)^\alpha,
\end{equation}
where $\epsilon$, $M_{\rm host}$ and $M_{\rm sat}$ are the circularity of the satellite's orbit, mass of host and satellite haloes, and $t_{\rm dyn}$ is the dynamical timescale
\begin{equation}
t_{\rm dyn}=\frac{R_{\rm vir}}{V_{\rm vir}},
\end{equation}
with $R_{\rm vir}$ and $V_{\rm vir}$ being the virial radius and circular velocity, respectively.
The $\alpha$ parameter \replaced{controls}{is the only fitting parameter that controls} the merger timescale of satellites.
Increasing $\alpha$ \added{on average} increases the number of satellite galaxies
and will enhance the overall amplitude of correlation functions as well as the FoG effect.
\replaced{Because $\alpha$ is set to be $1.5$ in \citet{Hong2016}
to visually find a good agreement with the correlation function of the SDSS volume-limited samples with $r$-band absolute magnitudes of ${\cal M}_r - 5\log h < -21$ and $-20$ \citep{Tempel2014},}{{Due to the limited computational resource, only three implementations for $\alpha=1.0, 1.5$, and $2.0$ are carried out.} In section \ref{section:results}, we will see that $\alpha = 1.5$ results in the best agreement between simulation and observation for the projected 2pCF of the volume-limited samples of galaxies with $r$-band absolute magnitudes of \deleted{$M_r < -21$ and} $M_r <-20$. \deleted{\citep{Zehavi2011}.}}
Therefore, we use $\alpha=1.5$ in this study as a fiducial model,
but also show some results and comparisons with other $\alpha$ values.

\subsection{The Horizon Run 4 Simulation}
The Horizon Run simulations \citep{Kim2009,Kim2015} are large cosmological $N$-body simulations run by \replaced{KIAS}{the Korea Institute for Advanced Study (KIAS)}.
To date, there are four realizations (Run 1, 2, 3, and 4) with different box sizes and particle numbers.
The Horizon Run 4 (HR4) %is $3150 h^{-1} {\rm Mpc}$ cubic in size and 
has evolved $6300^3$ particles with the mass of $3.0\times10^9 M_{\sun}$
in a $3150\ h^{-1}{\rm Mpc}$-long cubic box.
HR4 is $27$ times larger than the Multiverse \replaced{simulations}{Simulation},
allowing us to estimate the covariance matrix more accurately.
% and perform a systematics test (\S\ref{subsubsection:fiber}).
Adopted cosmological parameters are the same as those of the \replaced{$\Omega_m=0.26$}{$(\Omega_m, w) = (0.26, -1.0)$} case of
the Multiverse \replaced{simulations}{Simulation}.
While we use the Multiverse \replaced{simulations}{Simulation} to investigate the small-scale clustering property for different cosmological parameters,
we use the HR4 simulation to calculate the covariance matrix and to test systematics including the fiber collision effect.
The galaxy assignment was performed in an identical way to those of the Multiverse \replaced{simulations}{Simulation}.
To estimate the covariance matrix, we divide the HR4 simulation box into $5\times9\times5=405$ sub-cubes.
The choice of the number is because of the geometry of the SDSS Main Galaxy survey volume, whose length in one dimension is longer than those of the other two.
In each sub-cube, an origin is set and the galaxy positions are converted into $({\rm RA}, {\rm DEC}, z)$.
Then, the RSD effect is applied using the line-of-sight velocity of galaxies (see the next subsection).
We set $\alpha=1.5$ and the velocity bias parameter $b_v^s=1$ for the calculation of the covariance matrix.
The covariance matrix may be a function of these parameters, but we will ignore it.
The parameter fitting in section \ref{subsection:fitting} is performed using the covariance matrix obtained here, and all
error bars in the measurements of Figures \ref{figure:wrp}, \ref{figure:multipoles}, \ref{figure:FoG}, and \ref{figure:best} are the square root of the diagonal elements of the covariance matrix.

\subsection{The RSD and Velocity Bias}\label{subsection:vbias}
While the mock galaxy distribution is simulated in the real space, the observed galaxies clustering statistics come from the redshift-space distribution.
Thus, we need to apply the RSD effect to the simulation data.

The redshift-space distortion alters the apparent galaxy position \replaced{according to the line-of-sight component
of the peculiar velocities}{along the line of sight due to the peculiar motion in the radial direction} \citep{Hamilton1998}.
As we take the third axis of the simulation as the line-of-sight direction,
the positions are modified as
\begin{equation}
x_3^g \mapsto x_3^g + v_3^g/aH
\end{equation}
where ${\bf x^g} = (x_1^g,x_2^g,x_3^g)$ and ${\bf v^g}=(v_1^g,v_2^g,v_3^g)$ are
the comoving position and \added{peculiar} velocity of a galaxy, and 
%$a$ and 
$H$ is
%the scale factor and 
the Hubble parameter
at redshift $z$, respectively.
The periodic boundary condition is applied if the modified position exceeds the boundary of the simulation box.
For the first term of the right-hand side, we use the MBP positions as a proxy of galaxy positions.

\begin{figure}
\plotone{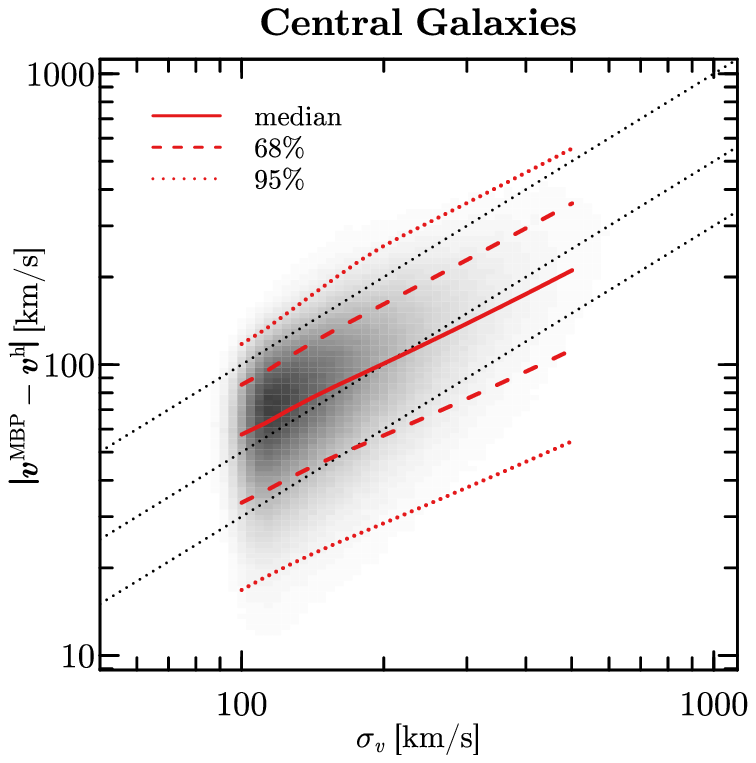}
\vspace{0.5cm}
\plotone{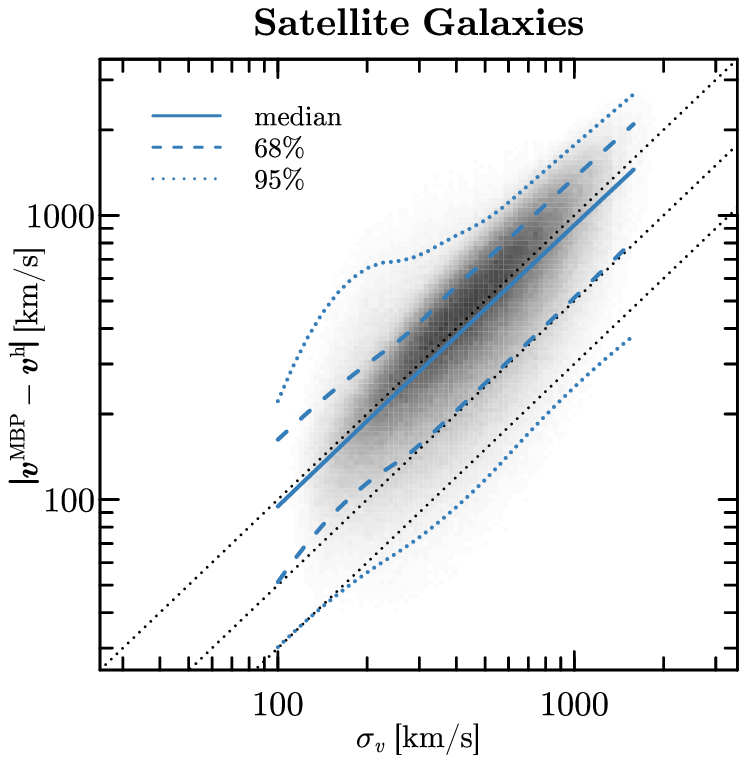}
\caption{%The comparison between the DM velocity dispersion and the MBP velocity in the halo frame.
The MBP's velocity in the halo frame versus the velocity dispersion of the DM halo particles.
The top panel is for central galaxies and bottom panel for satellite galaxies.
The colored lines indicate the median and the $68\%$ and $95\%$ percentile ranges.
\deleted{The solid lines show medians within each bin.}
The black dotted lines correspond to \replaced{$y=1.0x$, $0.5x$, and $0.3x$}{$|\mathbf{v^{\rm MBP}}-\mathbf{v^{\rm h}}|/\sigma_v = 1, 0.5$, and $0.3$}.
The model parameters are fixed to be $\Omega_m=0.26$ and $\alpha=1.5$.
}\label{figure:vbias}
\end{figure}

Recently it has been argued, based on the observations and simulations \citep{Munari2013,Wu2013,Guo2015,Ye2017},
that the galaxy velocity distribution may not be the same as that of DM inside haloes.
\citet{Guo2015} found that the speed of satellite galaxies inside haloes was lower
(typically $\sim 80\%$) than the velocity dispersion of the DM, $\sigma_v$.
Possible origins of such discrepancy include statistical bias, dynamical friction, galaxy interactions, and hydrodynamic effects.
Also, central galaxies may not be at rest at halo centers,
with the velocity dispersion of $\sim 0.3\sigma_v$.
Given that, we parameterize the satellite velocity bias by a single parameter $b_v^s$:

\begin{equation}\label{equation:velocitybias}
{\bf v^g}-{\bf v^h}=b_v^s({\bf v^{MBP}}-{\bf v^h}),
\end{equation}
where ${\bf v^{MBP}}$ and ${\bf v^h}$ are the MBP velocity of the galaxy and the host halo velocity, respectively.
The host halo velocity is defined as the average velocity of the member particles.
Equation (\ref{equation:velocitybias}) means that, in the rest frame of the hosting halo, the velocity of the visible part of galaxies is different \deleted{(usually slower)} from that of the \replaced{total galaxy mass}{all the matter} (represented by MBPs) by a factor of $b_v^s$.
Note that our definition of the velocity bias is different from that of \cite{Guo2015}.
They refer to $\alpha_v^s$ as the RMS velocity of satellites relative to the velocity dispersion of DM of their hosts: $\braket{|{\bf v^g}-{\bf v^h}|}=\alpha_v^s \sigma_v$
and use $\alpha_v^s$ to fit to the SDSS volume-limited sample.
Therefore, their $\alpha_v^s$ includes all factors that cause the velocity difference
between the baryonic component of galaxies and DM inside haloes.
Some of the factors for the velocity bias are hydrodynamic, and others gravitational.
Because the \replaced{MBP galaxy-halo}{MBP-galaxy} assignment approach naturally includes all gravitational effect,
the difference between $|\bf{v^{\rm MBP}}-\bf{v^{h}}|$ and $\sigma_v$ will reflect these effects.
As in Equation (\ref{equation:velocitybias}), our $b_v^s$ is defined as the difference between the visible component and \replaced{the total mass of a galaxy}{all the matter of the galaxy} represented by MBPs inside halo;
hence, $b_v^s$ will
indicate only the baryonic effects that cause the velocity bias.
Combining $\alpha_v^s$ and $b_v^s$ will tell us
to what degree each origin contributes to the velocity bias.
Figure~\ref{figure:vbias} shows the relation between $|\bf{v^{\rm MBP}}-\bf{v^{h}}|$ of centrals or satellites and $\sigma_v$ of DM haloes in the case of $\Omega_m=0.26$.
The colored lines show central $68\%$ and $95\%$ percentile intervals, which are obtained by quantile regression using B-splines \citep{Ng2015}.
The galaxy density is set to be similar to the observation data which we use.
The median of $|\bf{v^{\rm MBP}}-\bf{v^{h}}|/\sigma_v$ is $0.94$ for satellite galaxies;
the center of mass of satellite galaxies moves slightly slower than the velocity dispersion of DM inside the hosting halo.

One might wonder that the trajectories of MBPs and galaxies may diverge (i.e., the position of a galaxy in the next time step would be inconsistent with the corresponding MBP) if MBPs and galaxies have
different velocities.
Ideally, if the MBP represents the galaxy position and velocity correctly over cosmic time,
$b_v$ has to be one.
Our logic behind Equation (\ref{equation:velocitybias}) is that
we try to absorb the secondary effects
which may cause the velocity difference between the $N$-body simulation and the real observation, in response to the results of previous studies.
Although we expect that $b_v$ should be close to $1$ even if such effects are present,
a significant deviation from $b_v\sim1$, if detected, would indicate an incompleteness of using $N$-body simulations to fit the observational data on the small scales.

Considering that the central galaxies have spent a relatively longer time inside clusters and should be better relaxed ($\bf{v^g} \sim \bf{v^{MBP}}$),
the MBP velocity would be a good representative of the velocity of the  baryonic part of the central galaxy.
\added{Therefore, instead of making further sophistication, we use the velocity of the central MBP as the central galaxy velocity for most of our paper.}
As seen in Figure~\ref{figure:vbias}, the MBP velocity is in the range of $30\%$ to $50\%$ of $\sigma_v$.
This compares with the estimate on $\alpha_v^s\sim0.3$ in \citet{Guo2015}.
In an appendix, we will present the result obtained by modifying the MBP velocity for central galaxies.
Also, note that the velocity bias can be a function of galaxy property such as age and mass,
and studying the dependence of velocity bias in detail would help us to understand
the dynamical aspects of the evolution of galaxies, but we will only use a single parameter $b_v^s$ in this work.

In summary, our model parameters are
\begin{itemize}
\item matter density parameter: $0.15<\Omega_m<0.37$.
\item \replaced{halo to galaxy connection}{merger-time-scale parameter}: $\alpha=1.5$\replaced{, and}{.}
\item satellite velocity bias: $0.3<b_v^s<1.7$.
\end{itemize}

\section{Measurements}\label{section:measurements}

\subsection{Multipole Moments of the Correlation Function}
As a statistical quantity of the redshift-space clustering, we use the multipole moments of the correlation function.
First, the two-point correlation function is given by the Landy-Szalay estimator \citep{Landy1993},
\begin{equation}
\xi(\mathbf{s}) = \frac{DD-2DR+RR}{RR},
\end{equation}
where $DD$, $DR$, and $RR$ are the counts of galaxy-galaxy, galaxy-random, and random-random pairs, respectively.
The vector $\mathbf{s}$ can be $\mathbf{s} = (r_p,r_\pi)$ or $(s, \mu)$,
where $r_p$ and $r_\pi$ are the transverse and parallel components of the separation of galaxy pairs
while $s=|\mathbf{s}|$ and $\mu=r_\pi/s$. 

The multipole moments are calculated as
\begin{equation}
\xi_l(s)=\frac{2l+1}{2}\int^{1}_{-1} \xi(s,\mu)L_l(\mu)d\mu
\end{equation}
where $L_l(\mu)$ is the Legendre polynomial of $l$-th degree.
Because the moments of odd numbers vanish due to the symmetry\footnote{The relativistic effect can cause asymmetry by breaking the symmetry along the line of sight \citep{Alam2017}, but we do not consider it because the effect is much smaller than RSD.}
and higher order multipoles become less informative due to higher measurement noises,
we use only $l=0$ (monopole), $l=2$ (quadrupole), and $l=4$ (hexadecapole): $L_0=1$, $L_2(\mu)=\frac{1}{2}(3\mu^2-1)$, and $L_4(\mu)=\frac{1}{8}(35\mu^4-30\mu^2+3)$.
Because of the Kaiser effect, $\xi_0$ is enhanced  and $\xi_2$ becomes negative in redshift space
in scales larger than $\gtrsim 10 \ h^{-1}{\rm Mpc}$,
while the opposite holds at the cluster scales.
We use the bin size of  $\Delta \mu=0.05$ and \replaced{$9$}{$8$} logarithmic bins from \replaced{$s=0.98$}{{$s=1.43$}} to $30 \ h^{-1}{\rm Mpc}$.

\subsection{The Projected Correlation Function}
The projected correlation function is obtained by the integration along the line-of-sight,
\begin{equation}
w(r_p) = \int^{r_{\pi,{\rm max}}}_{-r_{\pi,{\rm max}}} \xi(r_p,r_\pi)dr_\pi.
\end{equation}
We set $r_{\pi,{\rm max}}=40 \ h^{-1}{\rm Mpc}$ and confirm that
larger $r_{\pi,{\rm max}}$ hardly changes $w(r_p)$.
The projected correlation function is a measure of clustering in real space,
because the line-of-sight projection eliminates the RSD effect,
whereas the multipole moments are the redshift-space quantities.
It will be shown that using the projected correlation function can break the degeneracy between the
\replaced{cosmological parameter}{cosmological matter density parameter} ($\Omega_m$), the \replaced{halo-galaxy relation}{merger-time-scale parameter} ($\alpha$) and the velocity bias ($b_v^s$) that are not fully broken by using multipoles only.

\subsection{The FoG Ratio}
\begin{figure}
\plotone{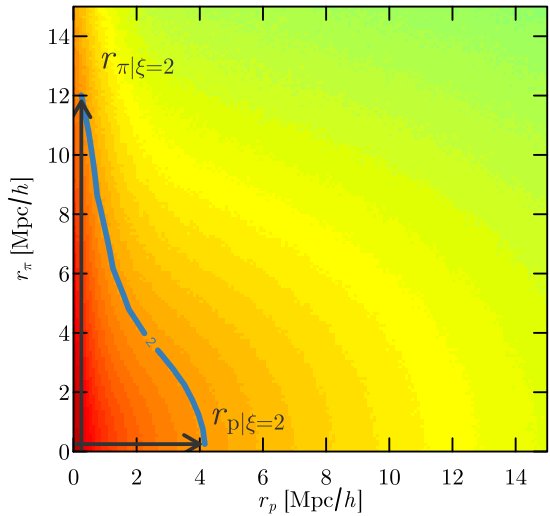}
\caption{The conceptual image of the FoG ratio.
The colored image is the redshift-space correlation function of galaxies
taken from the Multiverse simulation with $\Omega_m=0.26$.
The blue line shows the contour at the level of $\xi(r_p,r_\pi)=2$. 
The arrows corresponds to the numerator and denominator of Equation (\ref{equation:FoG}).
}\label{figure:image_R}
\end{figure}

Multipole moments of the correlation function are measures of the RSD effects,
but they are also affected by the change of the overall  clustering amplitude,
which can vary due to the cosmic variance and other systematics.
Thus, we try to extract a pure RSD information which is independent of the amplitude.
We use a measure of the strength of the RSD effects as follows,
\begin{equation}\label{equation:FoG}
R_{|\xi=3}=\frac{r_{\pi{|\xi=3}}}{r_{p{|\xi=3}}},
\end{equation}
which is the ratio of the separations along and across the line-of-sight from a point close to the origin to locations where the correlation function drops to $3$.
The ratio for different threshold values can be defined likewise.
The schematic image is given by Figure~\ref{figure:image_R}.
By taking a ratio, the cosmic variance in density fluctuations is expected to cancel out,
giving a clean measurement of the strength of the FoG effect.

We calculate the correlation functions for the Multiverse \replaced{simulations}{Simulation} and KIAS-VAGC catalogs,
covering $0.1<r_p<30 \ h^{-1}{\rm Mpc}$ and $0.1<r_\pi<30 \ h^{-1}{\rm Mpc}$ with $15\times15$ logarithmic bins.
Then, we take the fourth smallest bins ($\sim 0.4 \ h^{-1} {\rm Mpc}$), $\xi(0.4,r_\pi)$ and $\xi(r_p,0.4)$,
to locate the point at which the correlation function becomes a certain threshold value.
The scale $\sim 0.4 h^{-1} {\rm Mpc}$ is chosen to be sufficiently small to capture the FoG feature
while keeping statistical uncertainty small with enough pair counts.

\subsection{The Covariance Matrix}
The covariance matrix is necessary for evaluating the goodness of fit.
We use the mock galaxy catalogs created from the HR4 data,
which has a $3150^3 \ (h^{-1}{\rm Mpc})^3$ volume.
Using the $405$ mock catalogs from HR4, we have found that the distributions of our observables follow the Gaussian distribution. For each data point, we compared the distribution of mock values to the Gaussian distribution of the same mean and variance using the Kolmogorov-Smirnov test for the null hypothesis of the mocks following Gaussian. The resulting $p$-values are $0.4$--$0.9$, indicating no evidence for non-Gaussian distributions.
Therefore, we can use the standard $\chi^2$ statistics to evaluate the goodness of fit.

We adopt the $\chi^2$ statistics to constrain the model parameters,
\begin{equation}\label{equation:chi2}
\chi^2 = \left[{\mathbf X}^{\rm obs}-{\mathbf X}^{\rm th}({\mathbf \theta})\right]^T \mathbb{C}^{-1} \left[{\mathbf X}^{\rm obs}-{\mathbf X}^{\rm th}({\mathbf \theta})\right],
\end{equation}
where ${\mathbf X}$ is the data vector, $\mathbb{C}$ is the covariance matrix corresponding to ${\mathbf X}$, and the superscripts represent the observation
and the model prediction for parameters ${\mathbf \theta}=(\Omega_m, \alpha, b_v^s)$ respectively.
For example, if we use $\xi_0$ and $\xi_2$ for the fitting, ${\mathbf X}$ will be a vector of $8\times2=16$ elements and  \replaced{${\mathbf \rm C}$}{$\mathbb{C}$} be a $16\times16$ sized matrix.
We apply the correction of \cite{Hartlap2007} to the covariance matrix
to account for the underestimation of the covariance matrix due to
the finite number of realizations.
Because we use $405$ mock catalogs, the correction factor is
$1.02$ to $1.10$, depending on the size of the data vector.
Also, we multiply the covariance matrix by $(1+V_{\rm obs}/V_{\rm simu})=1.02$
to account for the uncertainty arising from the finite volume of the simulation box $V_{\rm simu}$ used to model the observation of volume $V_{\rm obs}$ \citep{Zheng2016}.

\section{Results}\label{section:results}
\subsection{The Correlation Functions}
\begin{figure}
\plotone{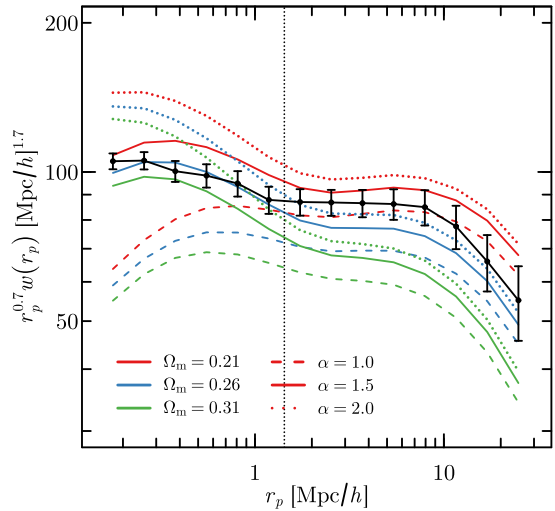}
\caption{The projected correlation function $w(r_p)$ for different $\alpha$ and $\Omega_m$.
The values are multiplied by $r_p^{0.7}$ for a visual purpose.
The black line represents the SDSS observation data,
while other colored lines are obtained from the Multiverse \replaced{simulations}{Simulation} for corresponding $\Omega_m$ values.
Different line types are for different merger time scale $\alpha$.
The dotted vertical line indicates the minimum scale for our fittings.
}\label{figure:wrp}
\end{figure}
The projected correlation function is shown in Figure~\ref{figure:wrp}.
Different colors correspond to different $\Omega_m$ while different line types to different $\alpha$.
Because we fix the overall \added{density perturbation} amplitude, $\sigma_8=0.79$,
increasing $\Omega_m$ shifts the matter-radiation equality,
resulting in weaker correlations at the scales which we are interested in.
An increase in $\alpha$ enhances the overall amplitude due to the increased
number of satellite galaxies.
The change is more drastic on small scales than large scales,
implying that the small-scale information is useful to discriminate different $\alpha$ scenarios,
in turn giving a better constraint on $\Omega_m$.
Also, it should be mentioned that the measurement error is small on smaller scales due to the larger number of pairs.
While $\alpha=1.0$ and $2.0$ fail to reproduce the observation,
the most probable value of $\alpha$ seems to be around $1.5$.
Note that $w(r_p)$ does not depend on $b_v^s$
because $w(r_p)$ is a real-space quantity and not affected by RSD.

\begin{figure}
\plotone{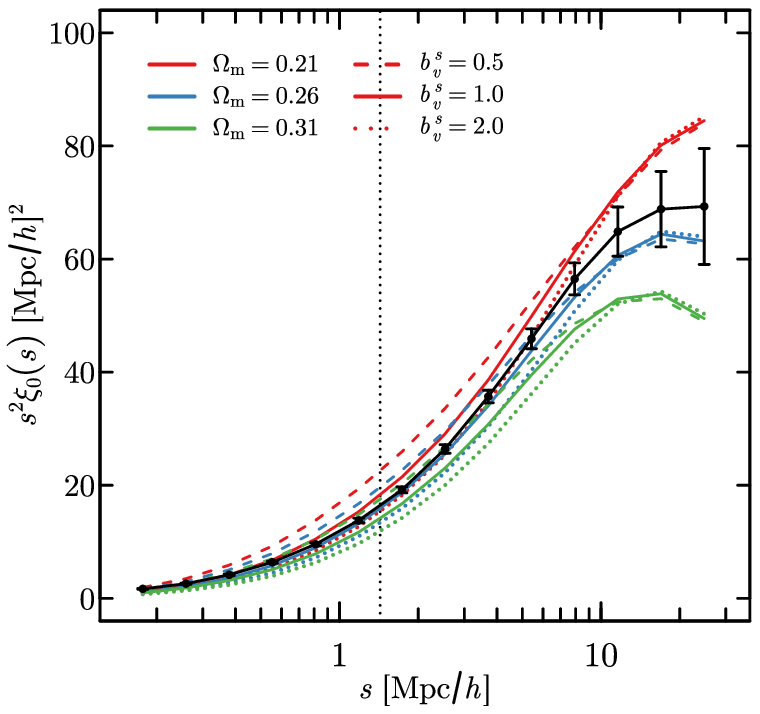}
\plotone{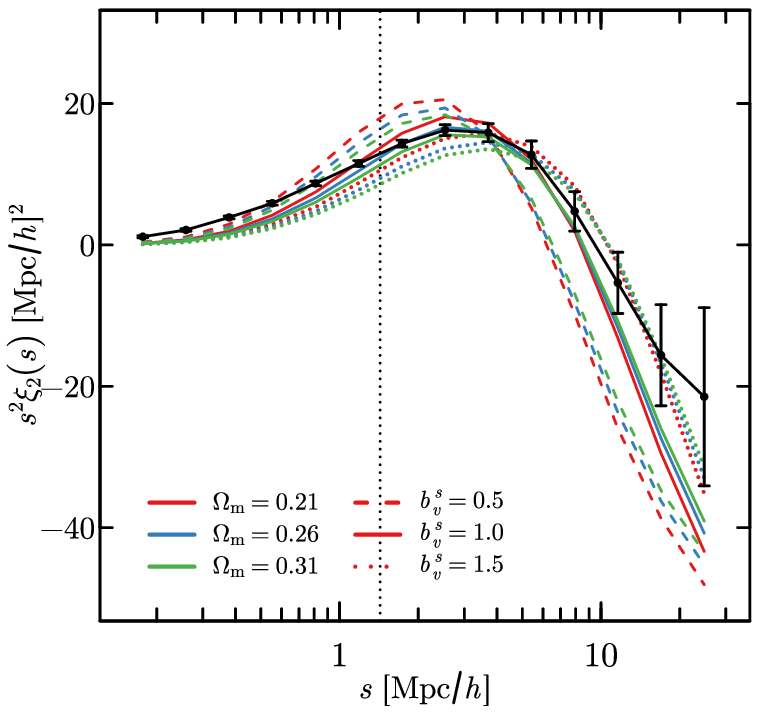}
\plotone{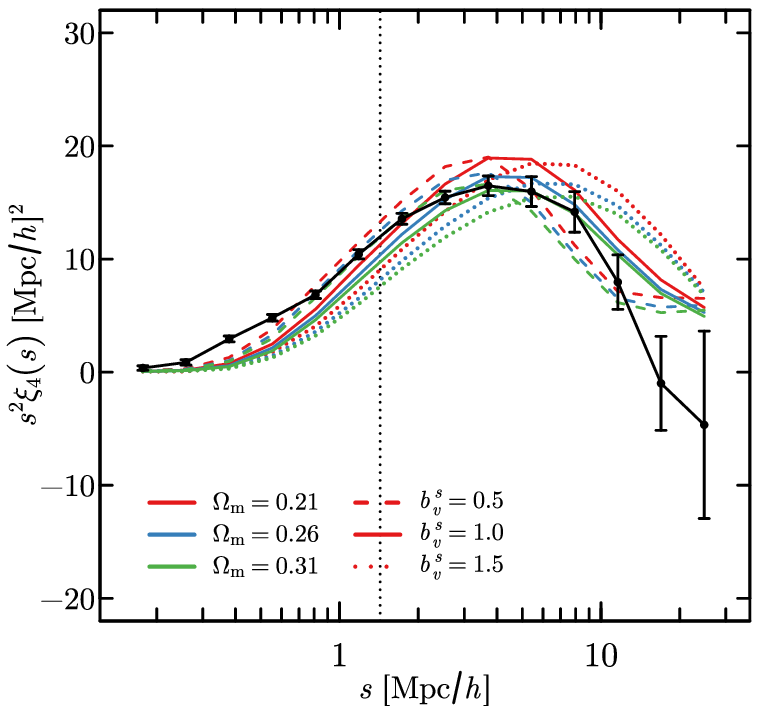}
\caption{The multipole moments $\xi_l$ ($l=0,2$, and $4$ for the top, middle, and bottom panels, respectively).
Black lines represent the observation data,
while other colored lines are obtained from the Multiverse \replaced{simulations}{Simulation} for corresponding $\Omega_m$ values.
Different line types are for different satellite velocity bias $b_v^s$.
The $\alpha$ parameter is fixed to be $1.5$.
}\label{figure:multipoles}
\end{figure}

Figure~\ref{figure:multipoles} shows the dependence of multipole moments
on $\Omega_m$ and $b_v^s$.
%Panels on different columns are for different $\alpha$ values,
Different panels are for different multipole moments.
The dependence of multipoles on $\Omega_m$ is complicated.
Both the Kaiser effect and FoG become stronger in a higher $\Omega_m$ universe
\citep{Feldman2003,Linder2005}
and thus $\xi_0$ should be suppressed (enhanced) at small (large) scales,
which is not the case at relatively large scales ($\sim 20 h^{-1}{\rm Mpc}$).
This contradiction is caused by the weaker real-space clustering for higher $\Omega_m$ as we saw in Figure~\ref{figure:wrp},
which is not fully compensated by the stronger Kaiser effect.
For $\xi_2$, the positive (negative) sign indicates the elongated (squashed) feature.
On larger scales, where the Kaiser effect dominates, $\xi_2<0$,
and on small scales, where the FoG does, $\xi_2>0$.
Again, a contradictory trend is seen in the middle panel of Figure~\ref{figure:multipoles},
which we attribute to the overall amplitude of the real-space clustering.
Another notable feature is the position of the peak of $\xi_2$.
If we increase $b_v^s$, the peak shifts toward larger $s$.
This is because large $b_v^s$ leads to a strong FoG effect,
increasing the transition scale from the FoG to Kaiser effect.
The position of the peak supports $b_v^s$ close to $1.0$.

\subsection{The FoG Ratio}
\begin{figure}\label{figure:FoG}
\plotone{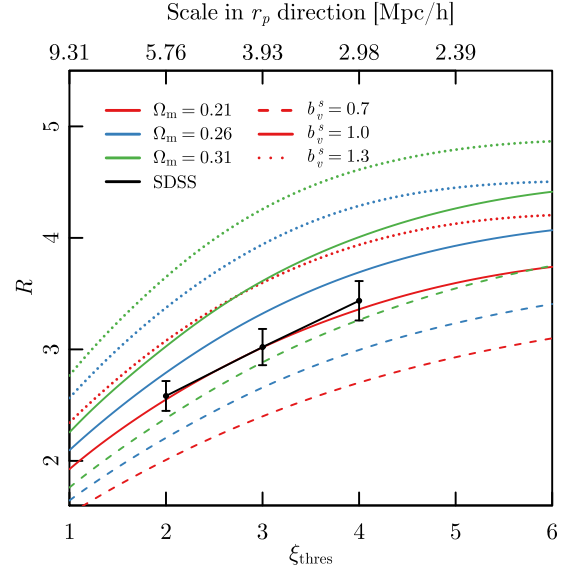}
\caption{The FoG ratio $R$ as a function of the correlation function threshold level.
The colored lines are obtained from the Multiverse \replaced{simulations}{Simulation} for different $\Omega_m$
while the black are from the SDSS observation.
Different line styles indicate different satellite velocity bias $b_v^s$.
The $\alpha$ parameter is fixed to be $1.5$.
For reference, approximate scales for corresponding threshold levels are shown
on the top axis.}\label{figure:FoG}
\end{figure}

Figure~\ref{figure:FoG} shows $R_{|\xi}$ for different $\Omega_m$ and $b_v^s$ as a function of the threshold value.
$R_{|\xi}$ is smaller for lower thresholds because lower thresholds correspond to larger scales where the FoG effect is \replaced{not}{less} dominant and the Kaiser effect becomes \added{more} effective,
which reduce $R_{|\xi}$.
A strong degeneracy between $\Omega_m$ and $b_v^s$ is seen.
\replaced{The higher $\Omega_m$ is, the more massive haloes there are and
the higher the virialized velocity is \citep{Marzke1995,Vikhlinin2009},
resulting in the stronger FoG effect.}{As $\Omega_m$ becomes higher, both the population of massive haloes and the virialized velocity become higher too \citep{Marzke1995, Vikhlinin2009}, which lead to the stronger FoG effect.}
Also, the higher velocity bias means the higher galaxy motion inside clusters and
thus the stronger FoG.
Within the error bars of the observation, both of $(\Omega_m, b_v^s)=(0.21,1.0)$ and $(0.31,0.7)$ reproduce the observed FoG ratio reasonably well.
If we had complete knowledge of $\alpha$ and $b_v^s$,
the FoG ratios would give a constraint of $\Delta \Omega_m \sim0.02$ with our data.
However, if we allow these to vary,
using only the FoG ratio is insufficient to obtain a meaningful constraint on $\Omega_m$.

\subsection{The Fitting}\label{subsection:fitting}
\begin{figure}
\plotone{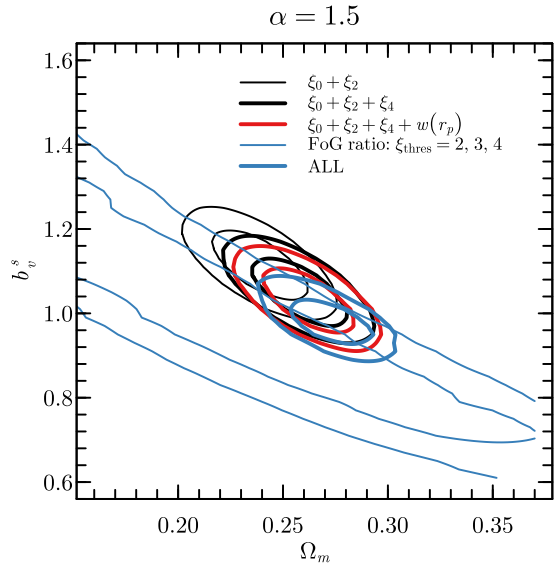}
%\plotone{figures/{logL_alpha2.0_0.98-30_full_mock}.eps}
\caption{\replaced{The $\chi^2$ for the fitting. The top panes shows that for the $\alpha=1.5$ case
while the bottom is for $\alpha=2.0$}{Constraints on $(\Omega_m, b_v^s)$ from the $\chi^2$ analysis for the case $\alpha = 1.5$.}
The contours show $68\%$ and $95\%$ confidence levels.
The constraints are obtained by using different combinations of the measurements
as given in different line types and colors.}
\label{figure:contour}
\end{figure}

Next, let us see how well the fittings work.
The peak position of the middle panel of Figure~\ref{figure:multipoles} tells us that
$b_v^s\sim1.0$ will give the best fit.
Then, we notice that $\Omega_m\sim0.26$ is preferred in the top panel by comparing the observation with the solid colored lines. 
%Also, $b_v^s\sim1.0$ is preferred at $5$--$10\  h^{-1}{\rm Mpc}$ of the bottom panel.
Figure~\ref{figure:contour} shows the probability distribution of $\Omega_m$ and $b_v^s$
based on the $\chi^2$ statistics for $\alpha=1.5$.
Because we only have three $\Omega_m$ realizations,
any statistical quantity ($\xi_l$, $w(r_p)$, and $R_{|\xi}$) for other $\Omega_m$ is obtained by interpolation.
Different lines correspond to what type(s) of information is(are) used to fit.
The bold red contour is the combined result obtained by fitting to the monopole, quadruple, and hexadecapole moments and the projected 2pCF.
The thin blue contour is from the FoG ratio and the bold blue is the combination of \replaced{the bold red and thin blue}{these two}.
Note that the projected 2pCF is the real-space quantity; therefore it cannot constrain $b_v^s$,
but can indirectly contribute to determining $b_v^s$ better by constraining $\Omega_m$.
All contours overlap one another in the $\alpha=1.5$ case.
This means that $\alpha=1.5$ can explain all measurements simultaneously, supporting the validity of our modeling of the galaxy clustering.
The best-fit values for the  $\xi_0+\xi_2+\xi_4+w(r_p)$ case are $(\Omega_m,b_v^s)=(0.262\pm0.014,1.032\pm0.051)$.
%Indeed, $\Omega_m\sim0.26$ and $b_v^s\sim1.0$ are suggested.
As seen in Figure~\ref{figure:contour}, including $\xi_4$ and $w(r_p)$ does not change the best-fit within statistical uncertainty nor
tighten the constraint significantly.
However, note that $w(r_p)$ has given a good constraint on
$\alpha$ as seen in \ref{figure:wrp}.
Adding the FoG ratio yields $(\Omega_m,b_v^s)=(0.272\pm0.013,0.982\pm0.040)$.

\begin{figure}
\plotone{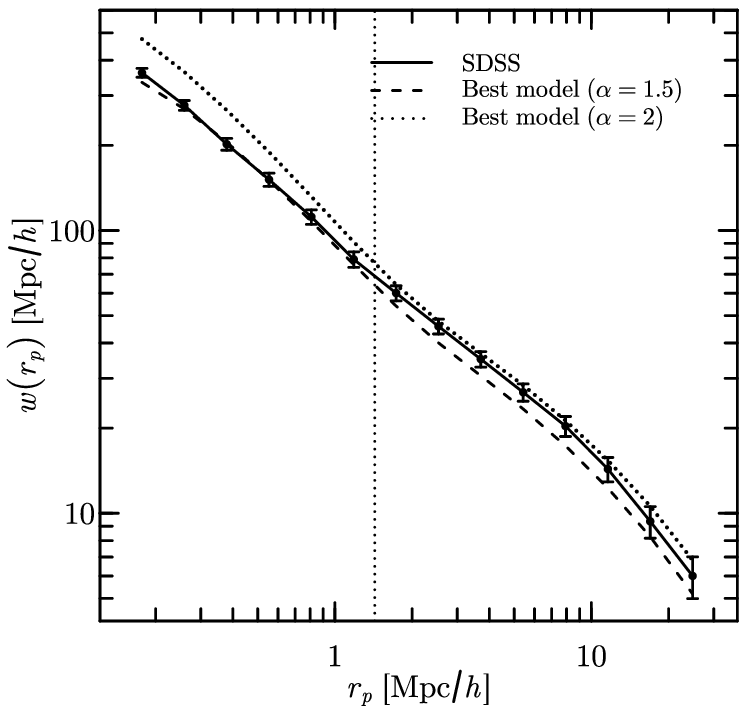}
\plotone{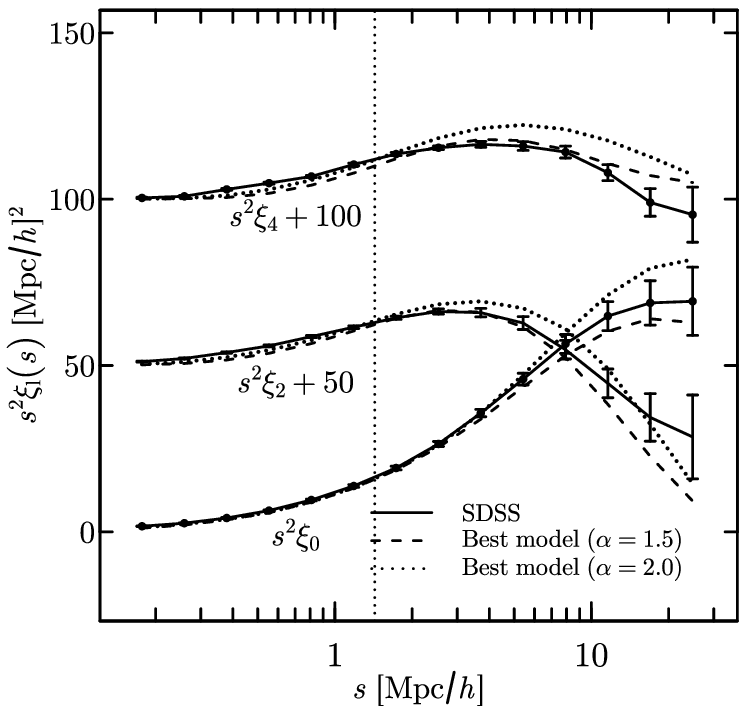}
\caption{The projected correlation function (top) and multipole moments (bottom).
The solid lines are the measurements from the SDSS.
The dashed line is the best model for $\alpha=1.5$,
obtained from the combination of $w(r_p)$, $\xi_0$, $\xi_2$, and $\xi_4$ (the red bold line in Figure~\ref{figure:contour}).
The dotted line is similar to the dashed line, but for $\alpha=2.0$ model.
For the log-likelihood of parameters for $\alpha=2.0$,
see Figure~\ref{figure:contour_2.0}.
The vertical lines show the minimum scale down to which we used
for the fittings.}\label{figure:best}
\end{figure}

\begin{figure}
\plotone{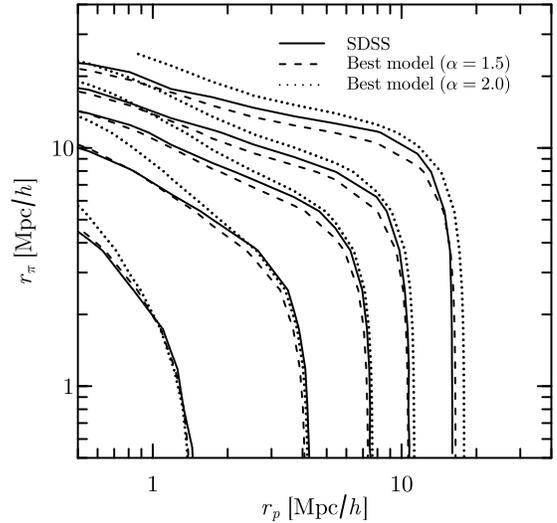}
\caption{The 2-D correlation function $\xi(r_p,r_\pi)$.
The contour levels are $6$, $2$, $1$, $0.6$, and $0.3$.
Axes are expressed in a logarithmic scale
to emphasize the small-scale part. 
The solid lines are obtained from the measurements of the SDSS data.
The dashed line is the best model for $\alpha=1.5$,
obtained from the combination of $w(r_p)$, $\xi_0$, $\xi_2$, and $\xi_4$ (the red bold line in Figure~\ref{figure:contour}).
The dotted line is similar to the dashed line, but for $\alpha=2.0$ model.
For the log-likelihood of parameters for $\alpha=2.0$,
see Figure~\ref{figure:contour_2.0}.
}\label{figure:xirppi}
\end{figure}

Figure~\ref{figure:best} gives a comparison between the observation and the best-fit models for $\alpha=1.5$ and $2.0$. 
The fitting procedure for $\alpha=2.0$ is identical to that for $\alpha=1.5$, except that we have used $\alpha=2.0$ in Equation (\ref{equation:Jiang08}) to produce the simulated galaxies in the Multiverse Simulation.
The top panel shows the projected 2pCF while the bottom the multipoles.
Except for a weak deviation ($\sim 1\sigma$) of at $s>10\ h^{-1}\rm{Mpc}$,
our $\alpha=1.5$ model reproduces $w(r_p), \xi_0$, and $\xi_2$ very well.
A moderate deviation is seen at $s\sim 1\ h^{-1}\rm{Mpc}$ of $\xi_4$, which mainly contributes to our $\chi^2$ (see appendix \ref{section:chi2} for detailed discussions).
Also, we can see a good fit of $w(r_p)$ at $r_p<1\ h^{-1}\rm{Mpc}$ in spite of our fitting range,
which also supports our galaxy model.
Figure~\ref{figure:xirppi} shows the 2-D correlation function for the best models.
We can see an excellent correspondence between the observation and our $\alpha=1.5$ model at small scales.
We see some deviation beyond $10\ h^{-1}\rm{Mpc}$, but this will be within $1\sigma$ uncertainty
as explained in Figure~\ref{figure:best}.

\section{Discussions}\label{section:discussions}
\subsection{Interpretation of $b_v^s$ and $\Omega_m$}
As we have seen in the previous section, the inferred velocity bias is $b_v^s=0.982\pm0.040$ for $\alpha=1.5$.
This value is slightly smaller than $1$, but the deviation is not significant for the size of the statistical error.
We will discuss how this result compares with other studies.
As we mentioned in \S\ref{subsection:vbias}, our parameter $b_v^s$ is different from the
definition used in the literature.
The velocity bias is usually defined as the velocity of visible part of galaxies
relative to the DM velocity dispersion inside their haloes.
Thus, it is the multiplication of two factors: the velocity bias of baryonic component of galaxies
(i.e., the observed galaxies) with respect to the whole galaxies \deleted{masses} (represented by MBPs),
and MBPs' velocity bias relative to the DM velocity dispersion.
The sources of the galaxy velocity bias are also separated into two classes.
The first one is related with the gravitational interactions such as dynamical friction, tidal stripping, and mergers. 
The other is baryonic effects including the star formation,
radiative cooling, feedback from stars/supernovae/AGNs and
the heat dissipation.
Given the fact that $N$-body simulations implement all gravitational effects,
the velocity of MBPs should reflect the velocity bias induced by gravitational interactions.
Since we have defined $b_v^s$ as the ratio between the velocity of the baryonic component of galaxies and MBPs, $b_v^s$ will indicate the velocity bias generated by hydrodynamic effects.
Figure~\ref{figure:vbias} shows that the median velocity of MBPs for satellite galaxies is $0.94$ of the DM velocity dispersion.
As a result, the total velocity bias amounts to $0.94 \times 0.982=0.93$ and
broadly consistent with the one obtained by \citet{Guo2015}.
Our results imply that the satellite velocity bias 
is attributed more to dynamical effects than baryonic effects.

\citet{Munari2013} ran $N$-body simulations with and without baryon cooling, star formation, and supernova and AGN feedbacks. 
Although it is not straightforward to compare their results with ours due to the different host halo mass focused on, their Figure 8 suggests $\sim10\%$ reduction of the velocity bias for the simulation with hydrodynamic effects.
This is because the star formation and radiative emission cool galaxies down,
forming a dense core and making galaxies resistant of tidal stripping.
Tidal stripping selectively disrupts slow-moving galaxies leading to higher mean galaxy velocity.
Thus, adding baryon cooling counteracts it and thereby reduces the averaged velocity.
\citet{Wu2013} investigated the effect of baryons on the galaxy velocity in their simulations, finding the reduction of galaxy velocity depending on the distance from the cluster center,
from $30\%$ (close to halo center) to $0\%$ (at virial radius of host haloes).
They suggest the reasoning for their result similarly to \citet{Munari2013},
but also mention the baryon dragging \citep{Puchwein2005}.
Considering that the fraction of satellite galaxies enclosed within the innermost region around the halo center is subdominant \citep{Watson2012},
the averaged reduction would be at most $10\%$.
\citet{Ye2017} found that the velocity bias depended on the ratio between the stellar mass and host halo mass,
implying that the velocity bias is mostly caused by the dynamical effects.
They argue that the dependence on stellar mass is a result of dynamical friction (a high-mass galaxy suffers from losing energy due to the two-body problem) and the dependence on host halo mass is related to the halo formation time (a high-mass halo is formed late, giving less time for dynamical effects to operate).
All of these studies indicate that the velocity reduction caused by baryonic physics will be less significant than that caused by dynamical effects, which agrees with our results.
\citet{Ye2017} also found that the velocity bias was a complicated function of other physical quantities, including age and color. Future studies can include investigating such dependence,
because knowing the detailed properties of the galaxy velocity bias would be useful for future surveys such as DESI and PFS, most of which apply color selections of galaxies to define the observation strategies.
They would also push forward our understanding of the kinematic perspectives of the galaxy formation and evolution.
For instance, we can classify galaxies by age, using the \replaced{SED}{spectral energy distribution} fitting technique, and measure the velocity bias through clustering measurements for each class.

Our constraint on $\Omega_m$ is $0.272\pm0.013$ when we use $\alpha=1.5$.
The value is consistent with the WMAP5 result ($\Omega_m=0.26$; \citealt{Dunkley2009}),
but lower than that of the Planck ($\Omega_m=0.31$; \citealt{Planck2018}).
In our simulation, the normalization of the power spectrum is set to give $\sigma_8=0.79$,
which is lower than the Planck results.
Thus, the correlation functions of our simulations are systematically weaker.
As we saw in Figures~\ref{figure:wrp} and \ref{figure:multipoles},
the correlation is stronger for lower $\Omega_m$,
which explains our $\Omega_m$ consistent with the WMAP5 rather than the Planck.
While we only run five simulations due to the large amount of resources required,
efficient methods of searching parameter space using $N$-body simulations
are being studied by several projects \citep{Nishimichi2018,DeRose2019}.
A more comprehensive study, including other cosmological parameters, would be beneficial.

\subsection{The Usefulness of FoG Ratio}
We have introduced a measure of the FoG strength as Equation (\ref{equation:FoG}).
As discussed in  \citet{Park2000} and \citet{Tinker2007},
%\citet{Tinker2007},
taking a ratio removes the dependence
on the overall amplitude of the real-space correlation function.
The clustering amplitude depends on cosmological parameters such as $\Omega_m$, $\sigma_8$, and the linear growth rate $f$,
\deleted{\textbf{PLEASE DEFINE $f$ HERE. (MT: Because the linear growth rate is referred to in the introduction, I have added the notation $f$ there.)}} which we usually wish to constrain,
but also \added{depends} on unwanted factors including cosmic variance and some sort of systematic errors.
The cosmological parameters inferred from only the multipole moments and
projected 2pCF can be contaminated by the latter factors. %even if the survey area is very large.
On the other hand, the FoG ratio is free from these uncertainties after division if these factors are universal. %\textbf{\replaced{universal, thus}{universal and, thus,}} can serve as a \textbf{\replaced{purer}{better}} probe of cosmological parameters than others.
As seen in the previous section, we should note that the constraining power is not strong because our FoG ratio uses the correlation function along $\mu=0$ and $1$ directions only.

\begin{figure}
\plotone{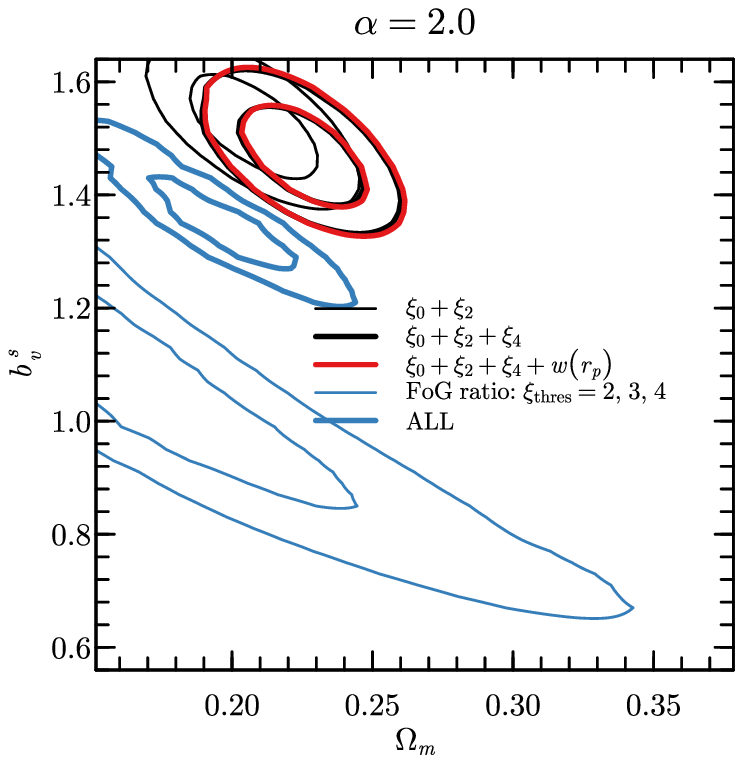}
\caption{\replaced{The $\chi^2$ for the fitting. The top panes shows that for $\alpha=2.0$.
The contours show $68\%$ and $95\%$ confidence levels.
The constraints are given from fitting to different combinations of the measurements
as given in different line types and colors.}{Same as the Figure~\ref{figure:contour} but for the case $\alpha = 2.0$.
Note that the contour from the FoG ratio disagrees with those from others.}}
\label{figure:contour_2.0}
\end{figure}

The FoG ratio depends on cosmological parameters differently from the multipoles and projected 2pCF.
Figure~\ref{figure:contour_2.0} shows the probability distribution of $(\Omega_m,b_v^s)$ for $\alpha=2.0$, which is to be compared with Figure~\ref{figure:contour}.
In each figure,
%Each of the $\alpha=1.5$ and $2.0$ case has its own preferred regions of 
combining $\xi_l$ and $w(r_p)$ gives preferred values of ($\Omega_m$, $b_v^s)$.
%$\Omega_m$ and $b_v^s$ when $\xi_l$ and $w(r_p)$ are used.
However, the contour from the FoG ratio disagree\added{s} with \replaced{that}{those} from the others in the $\alpha=2$ case,
which supports $\alpha=1.5$.
Noticeably, the contours from correlation functions and the FoG ratio shift toward different directions when the parameters are changed.

For the case of $\xi_l+w(r_p)$, increasing $\alpha$ results in lower $\Omega_m$ but higher $b_v^s$. The reason is as follows.
There are more satellite galaxies when $\alpha$ is increased, leading to higher amplitude of the correlation function.
On the other hand, increasing $\Omega_m$ decreases the amplitude of the galaxy 2pCF (see Figure~\ref{figure:wrp}).
This is because we fix $\sigma_8$, which means that the integration of the correlation over all scales remains the same.
Increasing $\Omega_m$ increases the amplitude on very large scales,
thus decreasing the correlation at the scales of our interest.
Therefore, $\alpha$ and $\Omega_m$ are anti-correlated.
In contrast, the positive correlation between $\alpha$ and $b_v^s$ stems from the amplitudes of $\xi_l$.
Considering the error bars of the observation, our constraints mainly come from $\sim1$--$3 \ h^{-1}\rm{Mpc}$.
At such scales, increasing $b_v^s$ reduces $\xi_l$ due to the enhanced FoG effect,
which compensates for the $\alpha$ increment.
The degeneracy between the three parameters are the result of the fact that $\xi_l$ and $w(r_p)$ depend on the overall clustering amplitude, unlike the FoG ratio.
Increasing both $\alpha$ and $b_v^s$, indeed, results in too strong FoG effect,
which can be seen in Figure~\ref{figure:xirppi}.

For the FoG ratio, on the other hand, an increase in $\alpha$ decreases both $\Omega_m$ and $b_v^s$.
This behavior is easily understood. 
Increasing $\alpha$ increases the number of satellite galaxies, resulting in stronger FoG.
In order to cancel this out, both $\Omega_m$ and $b_v^s$ need to be smaller.
Since the FoG ratio does not depend on the overall clustering amplitude,
how the parameters degenerate with one another is totally different from the 2pCFs.
Although we have mainly used $\alpha=1.5$ in our study,
the FoG ratio will tighten the constraints if we allow $\alpha$ to vary.
Changing $\alpha$ alters the galaxy-halo connection,
which is equivalent to changing the HOD parameters in that framework.
Therefore, the FoG ratio would also help to constrain the parameters in the HOD approach.
Another benefit of adding the FoG ratio would be that it gives a consistency check.
In the absence of the cosmic variance and systematic errors,
$w(r_p)$, $\xi_l$, and the FoG ratio overlap in the parameter space
if the correct model is chosen.
However, since $w(r_p)$ and $\xi_l$ are subject to the uncertainties of the clustering amplitude while the FoG ratio \added{is} not,
a discrepancy would be seen in the presence of systematic errors caused by the cosmic variance, observations, and data processing and analysis,
even if the employed fitting model were sufficiently accurate.

\section{Conclusions}\label{section:conclusions}
The small-scale galaxy clustering can provide a wealth of information about the cosmological model and galaxy-halo connection, owing to the availability of precise measurements.
In this study, we used the Multiverse \replaced{simulations \citep{Kim2015}}{Simulation \citep{Shin2017, Park2019}, the Horizon Run 4 Simulation \citep{Kim2015},} and a physically motivated galaxy assignment scheme \citep{Hong2016} to study the small-scale redshift-space clustering.
Specifically, we measured the projected correlation function $w(r_p)$ and
the multipole moments $\xi_l(s)$ of the correlation function from $1.4$ to $30\ h^{-1}\rm{Mpc}$
to examine their dependence on the matter density parameter $\Omega_m$ and the merger time scale parameter $\alpha$.
We also implemented the satellite velocity bias parameter $b_v^s$ to account for the
possible velocity difference between galaxies and dark matter inside haloes \citep{Munari2013,Wu2013,Guo2015,Ye2017}.
We have measured the correlation functions of a volume-limited sample from the KIAS-VAGC catalog \added{\citep{Choi2010}}, which is based on the SDSS DR7 spectroscopic data\added{,}
to compare with those of the Multiverse Simulation.
%constraining these parameters.
In the comparison, we have newly defined the strength of the FoG effect, $R_{|\xi}$,
which is free from the change in the overall amplitude of the correlation function
due to the cosmic variance and systematic errors.
We have found that $\alpha=1.5$ reproduce the observation well, with $(\Omega_m,b_v^s)=(0.272\pm0.013,0.982\pm0.040)$.
%which implies slightly slow motions of galaxies with respect to the MBPs.
While our $b_v^s$ broadly agreed with previous observational and simulation studies \citep{Munari2013,Wu2013,Guo2015,Ye2017},
$\Omega_m$ was smaller than the Planck results \citep{Planck2018},
which we attributed to the lower $\sigma_8$ that we assumed in the Multiverse simulations.
Considering that the velocity of MBPs for satellite galaxies are $0.93$ of that of dark matter,
the slow motions of galaxies relative to the dark matter velocity dispersion found by \cite{Guo2015} is mainly caused by dynamical effect rather than baryonic effects.
The FoG ratio was found to be useful to break the degeneracy between the parameters and can be used to check the consistency of the fit obtained by $w(r_p)$ and $\xi_l(s)$.

\begin{acknowledgments}
We thank Christophe Pichon for fruitful discussions on this work and Junsup Shim for useful comments on the paper.
\deleted{MT is grateful to the Korea Institute for Advanced Study for providing computing resources (KIAS Center for Advanced Computation Linux Cluster System).}
HP acknowledges the support by World Premier International Research Center Initiative (WPI), MEXT, Japan.
SEH was supported by Basic Science Research Program through the National Research Foundation of Korea funded by the Ministry of Education (2018\-R1\-A6\-A1\-A06\-024\-977).

Authors acknowledge the Korea Institute for Advanced Study for providing computing resources (KIAS Center for Advanced Computation Linux Cluster System). This work was supported by the Supercomputing Center/Korea Institute of Science and Technology Information, with supercomputing resources including technical support (KSC-2016-C3-0071) and the simulation data were transferred through a high-speed network provided by KREONET/GLORIAD.

Funding for the SDSS and SDSS-II has been provided by the Alfred P. Sloan Foundation, the Participating Institutions, the National Science Foundation, the US Department of Energy,
the National Aeronautics and Space Administration, the Japanese Monbukagakusho, the Max Planck Society, and the Higher Education Funding Council for England. 
The SDSS website is \url{http://www.sdss.org/}.

The SDSS is managed by the Astrophysical Research Consortium for the Participating Institutions. The Participating Institutions are the American Museum of Natural History,
Astrophysical Institute Potsdam, University of Basel, University of Cambridge, Case Western Reserve University, University of Chicago, Drexel University, Fermilab, the
Institute for Advanced Study, the Japan Participation Group, Johns Hopkins University, the Joint Institute for Nuclear Astrophysics, the Kavli Institute for Particle Astrophysics and Cosmology, the Korean Scientist Group, the Chinese Academy of Sciences (LAMOST), Los Alamos National Laboratory, Max Planck Institute for Astronomy (MPIA), the Max Planck
Institute for Astrophysics (MPA), New Mexico State University, Ohio State University, University of Pittsburgh, University of Portsmouth, Princeton University, the US Naval
Observatory, and the University of Washington.
\end{acknowledgments}

\clearpage
\appendix
\section{The $\chi^2$ statistics}\label{section:chi2}

Table~\ref{table:chi2} shows the minimum $\chi^2$ values and those per degree of freedom (d.o.f),
obtained from Equation (\ref{equation:chi2}) for different sets of measurements.
The best-fit $\chi^2/{\rm d.o.f}$ is $1.67$ for the case where we use all measurements,
which might be slightly high.
\replaced{%which is higher than the result of \citet{Guo2015} ($\chi^2/{\rm d.o.f} \sim 1$).
%There are several reasons for the difference.
}
{Here we give some possible reasons by discussing our statistical treatment, and suggest several ways to improving it.}

First, we used the covariance matrix estimated from the mock galaxy catalogs rather than jackknife resampling.
We have also measured the covariance matrix using the jackknife method
and found that the diagonal elements from the mocks are about half of
those from the jackknife method,
which means that the size of our error bars are smaller by a factor of $\sqrt{2}$.
Because doubling the covariance matrix halves $\chi^2$,
this partly describes the different $\chi^2$ values obtained by \citet{Guo2015} and us.
Then, why are the values of the covariance matrix from mock catalogs smaller than the jackknife resampling?
One reason is related to the inherent feature of the jackknife;
\citet{Norberg2009} demonstrated that the jackknife returned the error bars
accurately beyond $\sim 10\ h^{-1}\rm{Mpc}$
but significantly overestimated those below $10^{0.5}\ h^{-1}\rm{Mpc}$
for both $w(r_p)$ and $\xi(s)$,
where our constraints mainly come from.
Another possible reason is specific to our data;
it includes the Sloan Great Wall \citep{Gott2005},
which is centered at $z=0.08$.
The unusually huge structure may lead to the large region-to-region variance, enhancing the error bars from the jackknife.

\citet{Sinha2018} investigated the effect of the noise
from the limited number of mocks on the resulting $\chi^2$.
Due to the non-linearity of the inverse operation of the covariance matrix,
this kind of noise enters into a $\chi^2$ analysis in an unpredictable manner.
\citet{Sinha2018} provided a solution to use the principal component analysis to extract some eigenvectors with large signal-to-noise ratios,
and obtained smaller $\chi^2/{\rm d.o.f}$ values.

As obvious from Table~\ref{table:chi2},
$\xi_4$ contributes hugely to the large $\chi^2$ that we obtain.
However, Figure~\ref{figure:contour} shows that
the inclusion of $\xi_4$ does not improve the constraining power.
These facts might mean that the information of $\xi_4$ is already included
in the combination of $\xi_0$ and $\xi_2$
or our model is insufficient to reproduce up to $\xi_4$.
Improvements of our model can include allowing the central galaxy velocity bias parameter $b_v^c$ to change,
but we will only try $b_v^c=0$ in the next appendix and leave the detailed analysis to future works.

\begin{table}
\caption{The $\chi^2$ values \added
{(top) and $\chi^2$} per degree of freedom (d.o.f) \added{(bottom)} for the best-fit $(\Omega_m,b_v^s)$ cases for $\alpha=1.5$ and $2.0$.}
\begin{center}
\replaced{
\begin{tabular}{c|c|c|}
  & $\alpha=1.5$ & $\alpha=2.0$ \\
\hline
\hline
$\xi_0+\xi_2$ & 32.83/18 & 30.02/18 \\
$\xi_0+\xi_2+\xi_4$ & 101.00/27 & 119.26/27 \\
$\xi_0+\xi_2+\xi_4+w$ & 110.29/36 & 131.09/36 \\
$\xi_0+\xi_2+\xi_4+w+{\rm FoG}$ & 119.54/39 & 182.16/39 \\
\hline
\end{tabular}
}{
\begin{tabular}{l|c|c}
  & $\alpha=1.5$ & $\alpha=2.0$ \\
\hline
\hline
$w$ & 5.50 & 6.73 \\
$\xi_0+\xi_2$ & 13.08 & 25.27 \\
$\xi_0+\xi_2+\xi_4$ & 43.03 & 108.55 \\
$\xi_0+\xi_2+\xi_4+w$ & 47.45 & 114.56 \\
$\xi_0+\xi_2+\xi_4+w+{\rm FoG}$ & 55.03 & 171.02 \\
\hline \hline
$w$ & 0.92 & 1.12 \\
$\xi_0+\xi_2$ & 0.93 & 1.81 \\
$\xi_0+\xi_2+\xi_4$ & 1.96 & 4.93 \\
$\xi_0+\xi_2+\xi_4+w$ & 1.58 & 3.82 \\
$\xi_0+\xi_2+\xi_4+w+{\rm FoG}$ & 1.67 & 5.18 \\
\hline
\end{tabular}
}
\end{center}\label{table:chi2}
\end{table}

\section{The Model with Zero Central Galaxy Velocity Bias}
\begin{figure}
\plotone{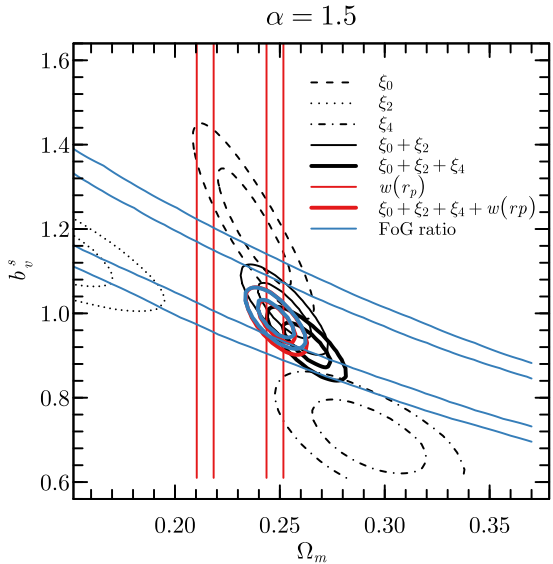}
\caption{Similar to Figure~\ref{figure:contour}, but $b_v^c=0$ is assumed.
Note that only the diagonal elements of the covariance matrix is used.}
\label{figure:contour_ac0}
\end{figure}

While we have assumed $\mathbf{v}^g \sim \mathbf{v}^{\rm MBP}$ for central galaxies in the main text,
we show the fitting result when the central velocity bias $b_v^c=0$,
i.e., the central galaxies are rest at the center of haloes ($b_v^c$ is defined similarly to Equation (\ref{equation:velocitybias})).
Figure~\ref{figure:contour_ac0} shows the $\chi^2$ contour obtained in the same manner as Figure~\ref{figure:contour}.
Although the final constraint (blue bold line) is apparently consistent with that in the main text,
it is simply a coincidence because each contour for $\xi_l$ does not overlap one another.

The shift of the FoG ratio (blue thin line) can be interpreted easily,
owing to the fact that the FoG ratio is a pure measurement of FoG.
The degree of the FoG effect is governed by the
quadratic sum of the velocities of central and satellite galaxies.
Therefore, $b_v^s$ has to be larger to compensate for
nullifying central galaxy velocities inside haloes.
On the other hand, understanding the shifts of the correlation functions
is not straightforward.
On small scales, the higher velocity bias takes the galaxy pairs to large separations in redshift space,
reducing the amplitudes of $\xi_l$.
Similarly to the FoG ratio, the decrease of the central galaxy velocity can be partly
canceled out by the increase of the satellite velocity bias.
However, not only the increase of $b_v^s$ but also the decrease in $\Omega_m$ can also increase the correlation amplitudes at such scales.
Furthermore, the data points from larger scales have to be fit simultaneously,
making the degeneracy of $(\Omega_m,b_v^c,b_v^s)$ complicated when we use the correlation functions.

Note that, only the diagonal elements of the covariance matrix is used to produce Figure~\ref{figure:contour_ac0}.
Due to the strong correlation and anticorrelation between
each bin of $\xi_l$ and $w(r_p)$,
the off-diagonal elements of $\mathbb{C}^{-1}$ are noisy.
If the theoretical model were reasonably correct, the contributions of off-diagonal elements to 
$\left[{\mathbf X}^{\rm obs}-{\mathbf X}^{\rm th}({\mathbf \theta})\right]^T \mathbb{C}^{-1}
\left[{\mathbf X}^{\rm obs}-{\mathbf X}^{\rm th}({\mathbf \theta})\right]$
would not affect the best-fit values significantly
because $|{\mathbf X}^{\rm obs}-{\mathbf X}^{\rm th}({\mathbf \theta})|$ is small.
While we have confirmed that the best-fit values in the main text were stable even if we use only the diagonal components,
we found that the blue bold line were afar from the rest contours when we used the full covariance matrix here.
This will imply that $a_c=0$ is not a good model,
probably giving unstable behaviors of the off-diagonal component of $\left[{\mathbf X}^{\rm obs}-{\mathbf X}^{\rm th}({\mathbf \theta})\right]^T \mathbb{C}^{-1}
\left[{\mathbf X}^{\rm obs}-{\mathbf X}^{\rm th}({\mathbf \theta})\right]$ caused by large $|{\mathbf X}^{\rm obs}-{\mathbf X}^{\rm th}({\mathbf \theta})|$,
which supports previous studies that claim the existence of the central galaxy velocity bias.

\section{The Fiber Collision Effect}\label{appendix:NNtest}
Due to the mechanical limitation of the SDSS spectroscopic instrument,
when a galaxy pair is separated by an angular separation less than $55''$,
only either one can be observed by a single run. 
This is called a fiber collision effect and
leads to systematics in the correlation function measurements (\citealt{Zehavi2002, Guo2012}).
Because the so-called nearest neighbor (NN) method is adopted in our study and it can cause systematic errors on the small scale measurements \citep{Reid2014}, we have tested the validity of it by simulating the fiber collision effect using the HR4 simulation data.
We use the concept of \cite{Guo2012} to model the fiber collision effect.\footnote{The public code is available at http://sdss4.shao.ac.cn/guoh/}
The angular FoF grouping is performed to the objects on the celestial plane with an linking length of $55''$.
Then, the objects are classified into three groups:
\begin{itemize}
\item  D1: galaxies which are isolated
\item  D2: galaxies which collide with one close galaxy (typically doublets)
\item  D2': galaxies which collide with more than one galaxy (typically the middle one of triplets)
\end{itemize}

We divide the HR4 galaxy catalog into $18$ boxes to create ``flux-limited" samples which correspond to the parent photometric catalog of the KIAS-VAGC. Specifically, we select the heaviest galaxies within the distance range,
which starts from $10\ {\rm Mpc}/h$ to $800\  {\rm Mpc}/h$ with a bin size of $10\ {\rm Mpc}/h$ to obtain the same number density.
Because the fiber collision occurs in the parent photometric catalog before any redshift and luminosity cut, we require much more distant galaxies, which leads to the small number of realizations (18) compared to the ones for the covariance matrix (405).

We apply this classification to the KIAS-VAGC parent catalog to estimate the fraction of fiber-allocated galaxies as a function of the population (D1, D2, or D2') and the number of plates covering the position of objects ($N_{\rm tile}$). 
Then, for each of the 18 realizations, the same classification code is run and ``observed" galaxies are determined according to these fractions.

\begin{figure}
\plotone{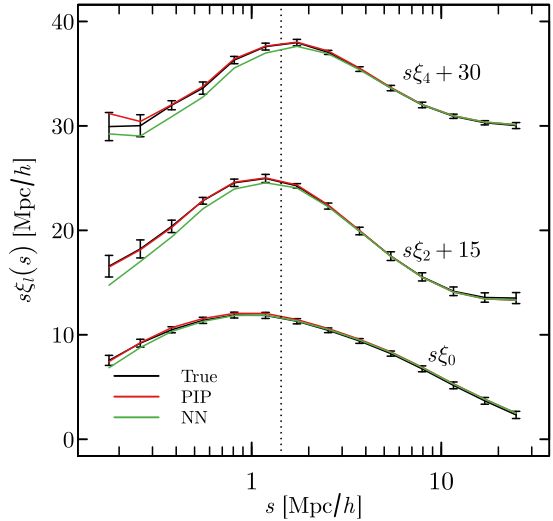}
\caption{The fiber collision effect on the multipole moments.
The black line is the fiducial data from the HR4 galaxy sample which we want to recover, while the red and green lines are obtained by simulating the PIP and NN scheme for the fiber collision correction, respectively.
The vertical line shows the minimum scale for our analysis in the main text. We adopt the NN scheme in this study.}\label{figure:fiber_multipoles}
\end{figure}

\begin{figure}
\plotone{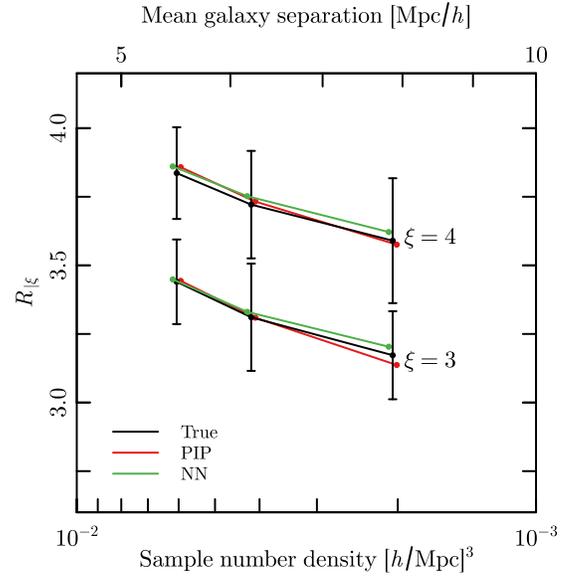}
\caption {The fiber collision effect on the FoG ratio.
We show two threshold levels, $\xi=3$ and $4$. 
The $x$-axis gives the sample number density. For our sample in the text, it is $6.1\times10^{-3} (h^{-1}{\rm Mpc})^3$ and corresponds to the leftmost points. 
The black line is the fiducial data from the HR4 galaxy sample which we want to recover, while the red and green lines are obtained by simulating the PIP and NN scheme for the fiber collision correction, respectively.
We adopt the NN scheme in this study.}\label{figure:collision}
\end{figure}

Then, we assign the nearest neighbor redshift to the ``unobserved" galaxies, apply the redshift and mass cut, and measure the correlation functions.
The comparison is given in Figure~\ref{figure:fiber_multipoles} and \ref{figure:collision}.
We also perform another fiber collision correction method based on the pairwise-inverse-probability weights (PIP; \cite{Bianchi2017}).
In this case, we repeat the selection process $1000$ times to create the logical array of length $N_{\rm bits}=1000$ for each galaxy, each elements of which is either 0 (unobserved) or 1 (observed). The correlation function is then measured using the pairwise weight given by Equation (14) of \cite{Bianchi2017}.

The PIP scheme is accurate over almost all scales, as it is an unbiased way of correcting for the missing observations.
The NN method is, however, still possible to use for our interested scales.
This result is different from the argument of \citet{Reid2014}, but can be explained by the difference of minimum scale probed
and the different collision scale in the comoving space ($<0.1 {\rm Mpc}/h$ and $<0.4 {\rm Mpc}/h$ for our and their works, respectively). 
While the higher redshift data such as BOSS and eBOSS would require the PIP method for small-scale clustering study, the NN method suffices for our study.


\end{CJK*}
\end{document}